\documentclass{jpp}

\usepackage{graphicx}
\usepackage{natbib}

\ifCUPmtlplainloaded \else
  \checkfont{eurm10}
  \iffontfound
    \IfFileExists{upmath.sty}
      {\typeout{^^JFound AMS Euler Roman fonts on the system,
                   using the 'upmath' package.^^J}%
       \usepackage{upmath}}
      {\typeout{^^JFound AMS Euler Roman fonts on the system, but you
                   dont seem to have the}%
       \typeout{'upmath' package installed. JPP.cls can take advantage
                 of these fonts, if you use 'upmath' package.^^J}%
      }
  \else
  \fi
\fi

\ifCUPmtlplainloaded \else
  \checkfont{msam10}
  \iffontfound
    \IfFileExists{amssymb.sty}
      {\typeout{^^JFound AMS Symbol fonts on the system, using the
                'amssymb' package.^^J}%
       \usepackage{amssymb}%
         
         \let\geq=\geqslant
      }{}
  \fi
\fi

\ifCUPmtlplainloaded \else
  \IfFileExists{amsbsy.sty}
    {\typeout{^^JFound the 'amsbsy' package on the system, using it.^^J}%
     \usepackage{amsbsy}}
    {\providecommand\boldsymbol[1]{\mbox{\boldmath $##1$}}}
\fi

\title[Z-pinches in space plasmas]{Role of Z-pinches in magnetic reconnection in space plasmas}

\author[V. Olshevsky, G. Lapenta, S. Markidis and A. Divin]%
{V\ls Y\ls A\ls C\ls H\ls E\ls S\ls L\ls A\ls V\ns O\ls L\ls S\ls H\ls E\ls V\ls S\ls K\ls Y$^{1,2}$%
  \thanks{Email address for correspondence: sya@mao.kiev.ua},\break
G\ls I\ls O\ls V\ls A\ls N\ls N\ls I\ns L\ls A\ls P\ls E\ls N\ls T\ls A$^1$\break
S\ls T\ls E\ls F\ls A\ls N\ls O\ns M\ls A\ls R\ls K\ls \ls I\ls D\ls I\ls S$^3$\break
\and A\ls N\ls D\ls R\ls E\ls Y\ns D\ls I\ls V\ls I\ls N$^4$}

\affiliation{
$^1$Centre for mathematical Plasma Astrophysics (CmPA), Department of Mathematics, KU Leuven, Celestijnenlaan 200B, bus 2400 B-3001 Leuven, Belgium\\[\affilskip]
$^2$Main Astronomical Observatory of NAS, 27 Akademika Zabolotnoho st., 03680, Kyiv, Ukraine\\
$^3$High Performance Computing and Visualization (HPCViz), KTH Royal Institute of Technology, Stockholm, Sweden\\
$^4$Swedish Institute of Space Physics, Disciplinary Domain of Science and Technology, Uppsala Division, SE-751 21, Uppsala, Sweden}

\pubyear{2014}
\volume{}
\pagerange{--}
\date{?; revised ?; accepted ?. - To be entered by editorial office}
\begin{document}

\maketitle

\begin{abstract}
A widely accepted scenario of magnetic reconnection in collisionless space plasmas is the breakage of magnetic field lines in X-points.
In laboratory, reconnection is commonly studied in pinches, current channels embedded into twisted magnetic fields.
No model of magnetic reconnection in space plasmas considers both null-points and pinches as peers.
We have performed a particle-in-cell simulation of magnetic reconnection in a three-dimensional configuration where null-points are present initially, and Z-pinches are formed during the simulation along the lines of spiral null-points.
The non-spiral null-points are more stable than spiral ones, and no substantial energy dissipation is associated with them.
On the contrary, turbulent magnetic reconnection in the pinches causes the magnetic energy to decay at a rate of $\sim1.5$\% per ion gyro period.
Dissipation in similar structures is a likely scenario in space plasmas with large fraction of spiral null-points.
\end{abstract}

\section{Introduction}
Magnetic reconnection is the main mechanism that causes the fast release of magnetic energy in space and laboratory plasmas.
The dissipated energy is transformed into heating, acceleration of particle jets, and generation of different plasma waves.
It is clear that reconnection in nature is essentially three-dimensional (3D), which was found in both space \citep{Xiao:etal:2006NatPh} and laboratory \citep{Intrator:etal:2009NatPh} plasmas.
Because classical collisions in space plasmas are often too weak to drive reconnection, non-ideal kinetic effects have to be considered, which limits the theoretical studies to rather simple two-dimensional geometries. 
In laboratory, the interest to magnetic reconnection is often motivated by the problem of plasma confinement with magnetic field, and specific laboratory experiments were built to study the reconnection itself \citep{Egedal:etal:2000RScI,Yamada:etal:2006PhPl,Furno:etal:2007PhPl}.
In the linear or toroidal configurations, much work has been done on energy dissipation in current channels surrounded by magnetic field, pinches \citep{Freidberg:2008,Yamada:etal:2010RvMP}.
The classic Sweet-Parker \citep{Parker:1957JGR,Sweet:1958IAUS} and Petschek \citep{Petschek:1964} paradigms of magnetic reconnection in X-points or neutral lines have been established fifty years ago, and many aspects of them are still used for interpretation of spacecraft measurements and simulations.

In the classical scenario, magnetic field lines of opposite polarity approach each other in the vicinity of an X-type magnetic null-point (or neutral line), where they break and reconnect.
The configurations that realize such scenario, are well studied in magnetohydrodynamic (MHD) and kinetic approaches \citep{Priest:Forbes:2000,Biskamp:2000}, and different reconnection regimes are classified \citep{Greene:1988,Lau:Finn:1990,Priest:Pontin:2009PhPl}.
Numerical simulations of magnetic reconnection in null-points in astrophysical and space plasmas were performed with MHD \citep{Galsgaard:Nordlund:1997JGR,Galsgaard:Pontin:2011b} and kinetic \citep{Baumann:Nordlund:2012ApJ} codes.
The test-particle studies by \citet{Dalla:Browning:2005,Stanier:etal:2012} found that null-points were able to accelerate particles to the energies, observed in solar flares.

A number of studies interpret spacecraft measurements during energetic events as null-point reconnection \citep{Xiao:etal:2006NatPh,Retino:etal:2007}. 
Alternative scenarios of magnetic reconnection in turbulent plasmas were proposed by \citet{Che:etal:2011Natur,Daughton:etal:2011NatPh} based on numerical simulations with topologies resembling X-lines.
Recently emerging is a scenario of reconnection in turbulence, reported in MHD \citep{Servidio:etal:2009PRL}, PIC \citep{Karimabadi:etal:2013PhPl} simulations, and in solar wind measurements \citep{Osman:etal:2014PRL}.
However, the major question remains unanswered: could these reconnection scenarios produce the observed high dissipation rates of magnetic field energy, or there are other mechanisms?

To reveal the dominant agent of magnetic energy dissipation in space plasmas, we have performed a fully kinetic electromagnetic particle-in-cell (PIC) simulation of a plasma configuration where null-points of different types where present, and pinches formed along neutral lines.
The simulation of a similar configuration in a small domain \citep{Olshevsky:etal:2013PhRvL} has shown a promising magnetic energy dissipation rate.
Here we report the simulations performed in a large domain, with extended analysis that allows to identify the features responsible for efficient energy release.

\section{Simulation setup}
\begin{figure}
\centerline{\includegraphics[width=\textwidth]{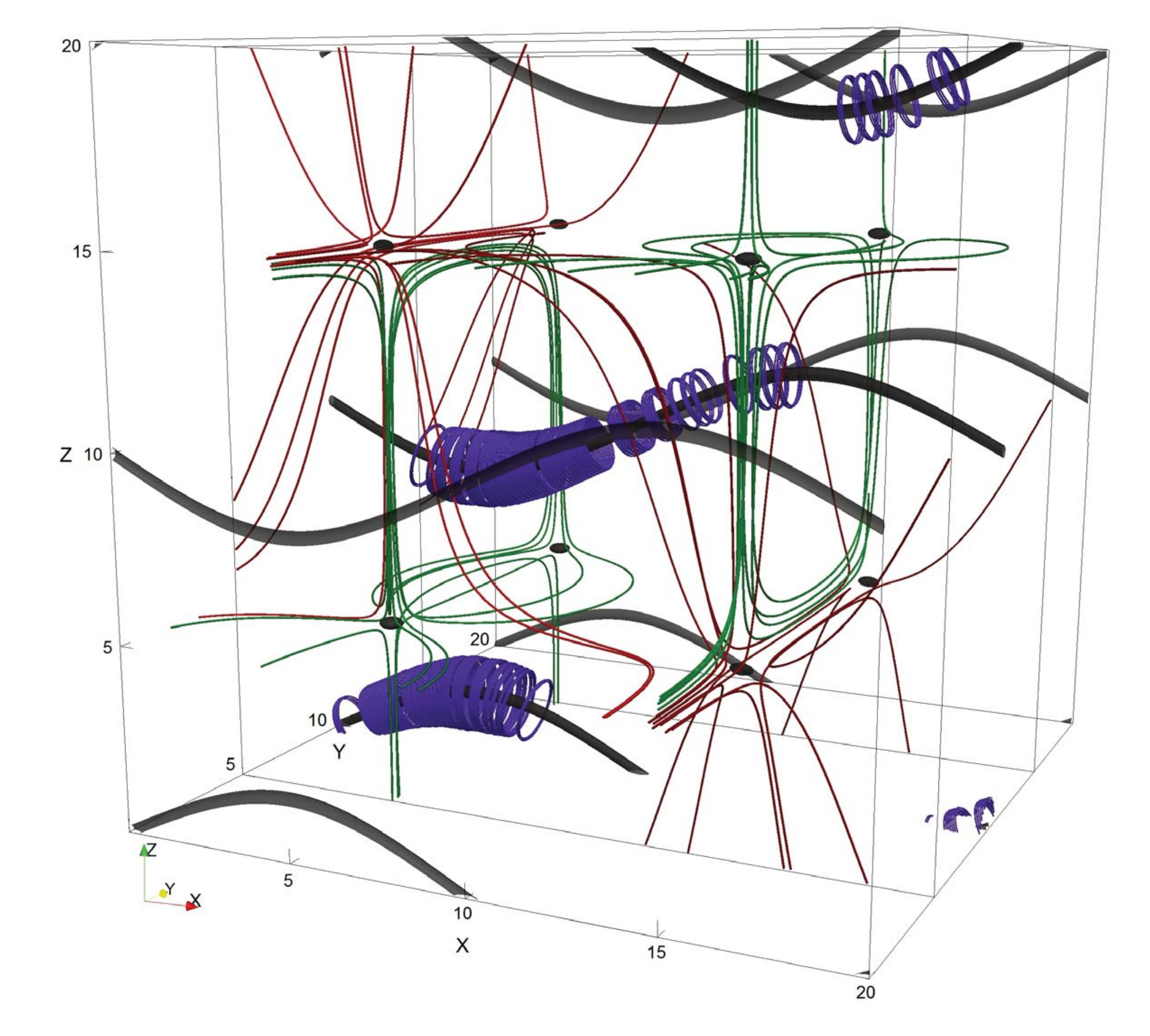}}
\caption{Initial magnetic field configuration.
Grey magnetic field isocontours at $B=0.1B_0$ depict the low magnetic field regions: separate null-points and null-lines.
The magnetic field topology of the null-points is illustrated by the green and red magnetic field lines.
In the vicinity of the null-lines the topology is different, as shown by the blue field lines that surround the null-lines lied in the $Y=10d_i$ plane.
The outline shows the slab of the domain spanning from $Y=5d_i$ to $Y=15d_i$ used to display the results.
}
\label{fig:1}
\end{figure}
To tackle the problem of realistic magnetic reconnection in 3D null-points, a carefully chosen initial setup is needed, 
which allows us to use the most advanced present-day physical model (PIC), but minimizes the influence of numerics on the results.
In PIC simulations, it implies use of periodic boundaries, because treatment of boundary conditions for particles is very complex.
It is easy to show that the minimum number of points where the magnetic field vanishes in a fully periodic 3D domain, is eight. 
To avoid further complications, we derived a divergence-free field configuration where eight null-points are uniformly spaced:
\begin{eqnarray*}
  B_x & = & -B_0\cos{\frac{2\pi x}{L_x}}\sin{\frac{2\pi y}{L_y}}, \qquad\qquad \\
  B_y & = & B_0\cos{\frac{2\pi y}{L_y}}\left( \sin{\frac{2\pi x}{L_x}} - 
        2 \sin{\frac{2\pi z}{L_z}} \right), \\
  B_z & = & 2 B_0\sin{\frac{2\pi y}{L_y}}\cos{\frac{2\pi z}{L_z}}, \qquad\qquad \\
\label{eq:initial}
\end{eqnarray*}
where $B_0=0.02$ is the magnetic field amplitude; $L_x$, $L_y$, and $L_z$ are the sizes of the simulation domain in the corresponding directions.
The condition $B=0$ holds in eight non-spiral null-points of A and B types following the classification of \citet{Lau:Finn:1990} (grey beads in Fig.~\ref{fig:1}), and along $9$ neutral (null) lines lied up in the planes $Y=0,\,L_y/2,\,L_y$ (grey channels surrounded by closed field lines in Fig.~\ref{fig:1}).
These lines are essentially the series of two-dimensional O-type null-points, hence are very unstable.
Z-pinches are created along these lines once the system starts to relax.

The remarkable symmetry of the initial configuration allows to display the results only in a part of the simulation domain for simplification.
The slab spanning from $Y=5d_i$ to $Y=15d_i$, used to represent our results, is indicated in Fig.~\ref{fig:1}.
Other regions are omitted, because they just repeat the topological features contained in the chosen slab.

The simulation of collisionless plasma was carried out using fully kinetic 
electromagnetic PIC code with implicit time stepping iPic3D \citep{markidis:etal:2010}.
The simulation domain representing a cubic box of size $20\times 20\times 20\, d_i$ (where $d_i=c/\omega_{pi}$ is ion inertial length) 
has $400^3$ cells with 27 particle of each specie per cell.
We considered two species: ions end electrons with mass ratio $m_i/m_e=25$.
Particles were initiated with a Maxwellian distribution with thermal speed in each dimension $u_{th,e}=0.02$ for electrons, and $u_{th,i}=0.0089$ for ions. 
This corresponds to the temperature ratio $T_i/T_e=5$, typical for the Earth magnetosheath plasma.
Given these parameters, the cell size is $0.25d_e$, sufficient for the implicit code to resolve all features important for energetics.
The time step is set to $0.15\omega_{pi}$, satisfying the finite-grid stability criterion; 
the total duration of the run was $52$ ion gyro periods $\Omega_{ci}^{-1}$.

We report the simulations initiated with uniform initial particle density and zero currents.
However, we have investigated the influence of both factors on the energetics of the system.
First, we ran simulations in which particle density was higher in the magnetized regions, and lower around null-points.
Such compensation for initial pressure imbalance just slows down the evolution, and doesn't change the major features of the energy evolution.
Influence of such features as initial density distribution or, e.g., ``guide'' magnetic field on the system are, in our opinion, the subject of a separate study.
Second, a simple estimate gives that the initial currents derived as $\mathbf{J}=\mathbf{\nabla}\times\mathbf{B}$ are by order of magnitude smaller than 
powerful currents established in the first phase along neutral lines, and should not influence the evolution.
For the sake of completeness we have initiated the simulation with initial currents given by the curl of magnetic field, and indeed no significant influence on the evolution was found;
these currents decay quickly after the relaxation starts.
We report the simulation with zero initial currents because we intend to compare our PIC simulations with MHD.

\section{Evolution}
\begin{figure}
\centerline{\includegraphics[width=\textwidth]{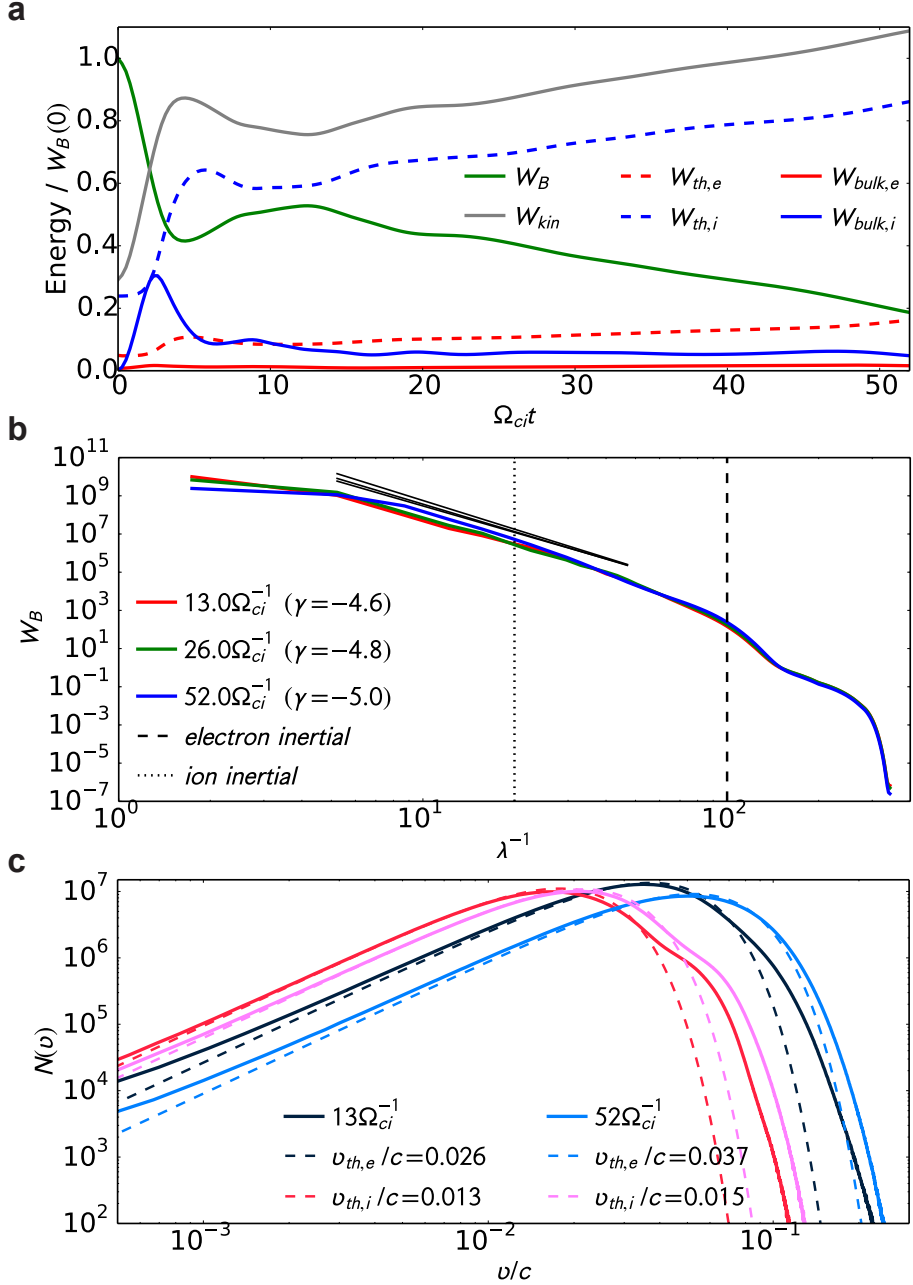}}
\caption{Domain-averaged quantities.
{\bf a}. Evolution of different components of energy.
The energy of magnetic field (green) drops off during the initial relaxation, then slightly rises, and steadily dissipates in the last phase of the simulation.
The dissipated magnetic energy is gained by particles (grey).
Powerful ion currents are excited in the beginning (blue solid), that dissipate in the second phase. 
The energy of electron bulk flows (red solid) is very small, however, in the last phase they are the dominant carriers of the currents.
In the turbulent reconnection phase, the dissipated energy dominantly goes to heat the ions (blue dashed) and electrons (red dashed).
{\bf b}. Evolution of the magnetic energy spectrum.
Initially, the energy is concentrated only in large-scale features, but within a few ion gyration times the cascade establishes (red) with exponential slope between ion and electron scales.
Two breaks of the spectrum are noticeable: a poorly resolved one at large scales, where, in principle, transition to MHD turbulence should happen, and another one at electron scales.
{\bf c}. Non-Maxwellian velocity distributions for ions (red) and electrons (blue) after the first phase (darker colors), and at the end of the simulation (lighter colors).
The distributions are computed over all particles, i.e., integrated over all pitch angles.
The dashed lines are the corresponding Maxwellian best-fits.
It is clear that suprathermal ions are created during the initial relaxation.
}
\label{fig:2}
\end{figure}
As noted above, our initial state is not in equilibrium, and the evolution starts immediately after the beginning of the simulation, driven by a large pressure imbalance:
initially magnetic field energy consitutes $77$\% of the total energy, resembling a low-beta space plasma with gas/magnetic pressure ratio $\beta\geq 0.013$. 
Although such initial assumption is contradictory, it may well correspond to a bifurcation state formed, e.g., after the propagation of a dipolarization front.
The evolution of the magnetic and particle energies in the simulation is shown in Fig.~\ref{fig:2}{\bf a} (the energy of the electric field is negligible).
The simulation proceeds in three distinct phases: rapid relaxation and formation of pinches, lasting $4\Omega_{ci}^{-1}$; reverse energy exchange, from $t=4\Omega_{ci}^{-1}$ to $t=13\Omega_{ci}^{-1}$; and stationary reconnection henceforth.
Magnetic and kinetic energies counterpart each other already at few ion cyclotron times, but the relaxation continues until almost 50\% of magnetic energy is dissipated.
In the second phase gas pressure dominates over magnetic pressure, and when the initial imbalance is fully compensated, the third phase of the evolution begins.
At the end of the simulation, magnetic energy constitutes only $15$\% of the total energy of the system: more than $80$\% of the initial magnetic energy is released to the particles within $50\,\Omega_{ci}^{-1}$.
The stationary reconnection phase begins when the magnetic energy is twice smaller than the initial value; $60$\% of this energy is released during this phase, giving the magnetic energy decay rate of $\sim 1.5$\% per ion gyro period.

In the first phase, due to the pressure imbalance, particles are pushed towards the regions of small magnetic field, and form current wires along the null-lines. 
As noted before, these currents are by an order of magnitude stronger than those given by the curl of magnetic field; this effect can not be observed in a single-fluid MHD simulation.
Ions are heavier than electrons, and carry almost all the bulk kinetic energy of current flows, as illustrated by a sharp peak in Fig.~\ref{fig:2}{\bf a}.
In this process, a substantial amount of particles is accelerated to suprathermal speeds, especially prominent in the ion velocity distribution in Fig.~\ref{fig:2}{\bf c}.
Later the ion currents decrease, the ion bulk energy dissipates, part of it goes back to electromagnetic field in the second phase of the evolution, and the currents are mostly carried by electrons.
Eventually, thermal energy dominates over the energy of bulk motions, and, due to the their higher mass, the ions gain by a factor of $\sqrt{m_i/m_e}$ more energy than the electrons.
Interestingly, the ion/electron temperature ratio holds with time, which was not the case in the simulation in a smaller domain \citep{Olshevsky:etal:2013PhRvL}.
The conditions under which the $T_i/T_e$ ratio substantially changes during the simulation, may be related to the initial gas/magnetic pressure ratio, and need further investigation.

Energy dissipation in turbulent plasmas happens at small, diffusive, scales, producing characteristic cascades in the spectrum.
In our simulation initial magnetic field energy is concentrated only in large-scale features.
After the simulation is initialized, and relaxation begins, magnetic energy is redistributed to smaller scales where it dissipates.
The characteristic shape of the spectrum, established within a few ion gyration times, has two noticeable breaks: at large (fluid) scales, and at electron scales. 
Between ion and electron scales the spectrum has a power-law shape with $\gamma\approx-4.6$.
This slope steepens with time, and reaches $\gamma\approx-5$ at the end of the simulation.
While there are indications that at sub-electron scales in solar wind $\gamma\approx-4$ \citep{Sahraoui:2013ApJ}, our values of $\gamma$ are too high for any 2D or 3D turbulence according to present knowledge \citep{Brandenburg:Lazarian:2013SSR}; they may be a consequence of the dominating, essentially 1D, processes in pinches. 

The Alfven crossing time in our simulation is $t_A=5\Omega_{ci}^{-1}$, hence the simulation is too short for a turbulent cascade to establish, and there is no external energy supply typical for the simulations of turbulence, therefore we describe the last phase of our simulations as ``relaxing turbulence''.
It is evident that magnetic reconnection plays an important role in our simulations (Fig.~\ref{fig:4}), but it is impossible, at this point, to distinguish the exact mechanism beyond energy conversion (see, e.g., \cite{Karimabadi:etal:2013SSRv} for the recent review on this subject).

The dissipation processes in the first phase are capable of accelerating particles to non-Maxwellian velocities (Fig.~\ref{fig:2}{\bf c}).
The power-law tails at high-energy parts of ion spectra in space plasmas are long known \citep{Montgomery:etal:1968JGR,Collier:etal:1996GeoRL}, the power-law exponent of $\sim-5$ is often detected in ion velocity distribution in space \citep{Gloecker:2003}.
During our simulation such tail is formed only for a narrow range of velocities (Fig.~\ref{fig:2}{\bf c}), and the exponent $\approx-4$ is not as steep as typically measured, e.g., in solar wind.
A simulation with non-Maxwellian initial particle distributions might be a better way to reproduce the observed power-law tails.
In the stationary reconnection phase the dissipation dominantly heats particles, and the suprathermal tails do not substantially change.

\section{Z-pinches}
\begin{figure}
\centerline{\includegraphics[width=\textwidth]{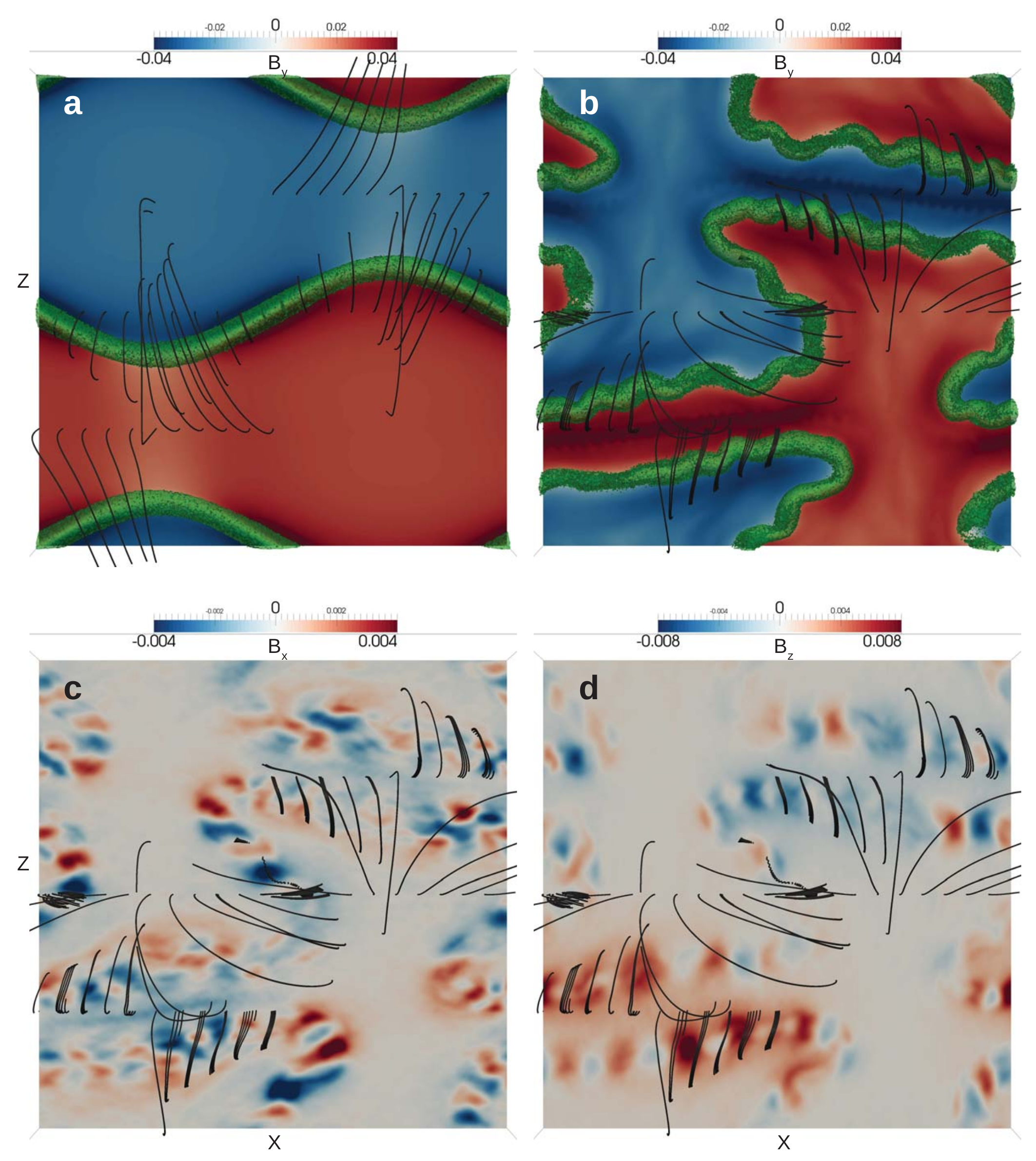}}
\caption{Magnetic field and Z-pinches in the $Y=10d_i$ plane.
Currents (green isocontour of current density $J=0.005$) in the null-lines are already established at $2.6\Omega_{ci}^{-1}$ ({\bf a}); in the middle of the run, at $26\Omega_{ci}^{-1}$ ({\bf b}) the pinches are already kinked.
At this time, the kinking has generated the in-plane magnetic fields $B_x$ ({\bf c}) and $B_z$ ({\bf d}). 
The complex structure of magnetic field is illustrated by the black field lines.
}
\label{fig:3}
\end{figure}
Two components of the magnetic field, $B_x$ and $B_z$ are zero in the $Y=10d_i$ plane in the initial configuration.
Where the third component, $B_y$, also turns to zero, the null-lines are formed (Fig.~\ref{fig:1}). 
These null-lines are essentially the series of the O-type null-points, surrounded by closed magnetic field lines.
Due to pressure imbalance, the particles move towards these regions of low $B$, and create current wires with radii $\sim d_i$ (Fig.~\ref{fig:3}{\bf a}).
These self-induced currents are by order of magnitude stronger than the currents derived from Ampere's law for the initial magnetic field configuration, and are efficiently confined by compressing pinch force, forming Z-pinches.

The Z-pinches are extremely unstable to various modes \citep{Kadomtsev:1966,Freidberg:1987}, including the kink mode, which is dominant in our simulation (Fig.~\ref{fig:3}{\bf b}).
The currents in the neghboring pinches are antiparallel, and the repulsion between them prevents their merging and disruption of the current system.
Kinking generates small amplitude mixed-polarity in-plane fields $B_x$ and $B_z$ around the null-lines (Fig.~\ref{fig:3}{\bf cd}).

\begin{figure}
\centerline{\includegraphics[width=\textwidth]{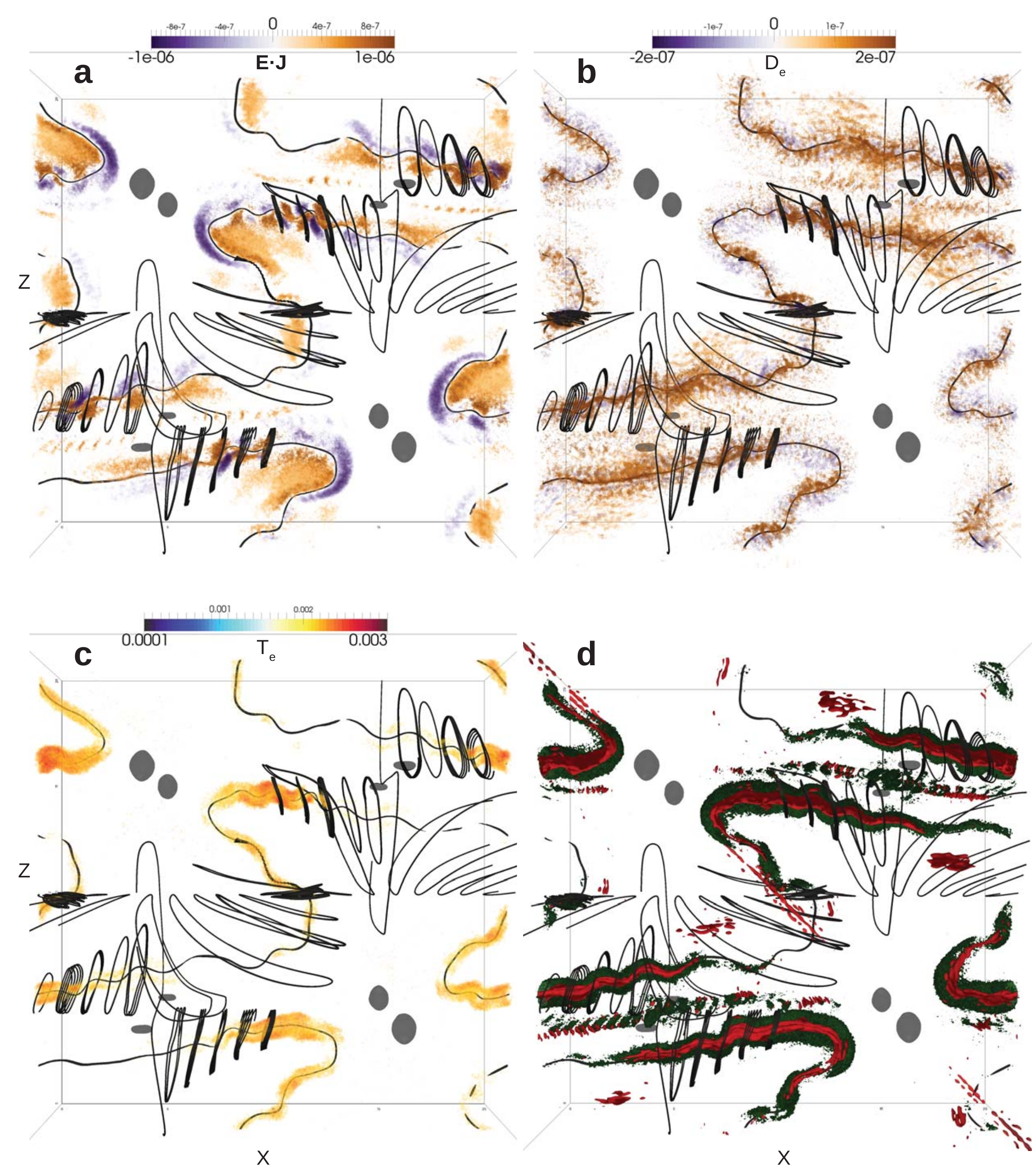}}
\caption{Indicators of magnetic reconnection.
{\bf a}. $\mathbf{E}\cdot\mathbf{J}$ can be both positive and negative around the pinches, depicting their oscillations.
{\bf b}. The electron dissipation measure $D_e$ shows the regions of intense particle heating.
{\bf c}. The electron temperature $T_e$ is high where the pinches are bended.
{\bf d}. Electrons are decoupled where the frozen-in condition is violated:
green is an isocontour of $\left|\mathbf{\upsilon}_{e,\perp}-\left[\mathbf{E}\times\mathbf{B}\right]/B^2\right| = 0.03c$ (decoupling);
red is an isocontour of $\left|\mathbf{B}/B\times\left[\mathbf{\nabla}\times\mathbf{S}\right]\right|=1.3\cdot10^{-3}$ (frozen-in violation).
Grey magnetic field isocontours at $B=0.1B_0$ depict the low magnetic field regions; grey magnetic field lines show the poloidal fields of the Z-pinches. 
}
\label{fig:4}
\end{figure}

We apply different approaches to indicate the features important for the magnetic energy dissipation and particle acceleration.
The work of the electromagnetic field on the plasma $\mathbf{E}\cdot\mathbf{J}$ is high around the Z-pinches, but zero in the current channels (Fig.~\ref{fig:4}{\bf a}).
Patches of negative and positive $\mathbf{E}\cdot\mathbf{J}$ follow the kinked shape of the Z-pinches, indicating the oscillations of the current wires, and changes of the particle drift velocity.
The total work of the fields on particles integrated over the domain is positive, as clearly shown in the energy evolution plot Fig.~\ref{fig:3}{\bf b}.
Similar mixed-sign distribution of $\mathbf{E}\cdot\mathbf{J}$ is observed in the simulations of \cite{Lapenta:etal:2014PhPl}, where also the integrated value is positive.

A recently proposed electron-frame dissipation measure is more adequate to represent the regions of intense heating \citep{Zenitani:etal:2011}.
Since the charge separation in our simulation is negligible, the dissipation measure can be estimated as 
$D_e = \mathbf{J}\cdot\left( \mathbf{E} - \left[\mathbf{\upsilon}_e\times\mathbf{B}\right] \right)$, where $\mathbf{\upsilon}_e$ is electron speed.
Vast regions of positive $D_e$ in Fig.~\ref{fig:4}{\bf b} co-align with the mixed-polarity patches of the in-plane magnetic field (Fig.~\ref{fig:4}{\bf c, d}), and clearly show that the energy is gained by particles inside and around the Z-pinches; much smaller volume is occupied by the regions of negative $D_e$.
The regions of intense heating along the current wires are complemented by the regions with the hottest plasma (Fig.~\ref{fig:4}{\bf c}), however no indication of energy dissipation is seen around the eight non-spiral null-points. 
High temperatures in the vicinity of current sheets, at proton scales, have been found in solar wind \citep{Osman:etal:2011ApJL}.

The formal criterion for magnetic reconnection, which means local breaking of conservation of the magnetic field topology \citep{Hesse:Schindler:1988JGR}, 
can be expressed as 
$\mathbf{\hat{b}} \times \left[ \mathbf{\nabla} \times \mathbf{S} \right] \ne 0$,
where $\mathbf{\hat{b}}=\mathbf{B}/B$ is the direction of magnetic field, 
and $\mathbf{S}=\left(\mathbf{E}\cdot\mathbf{B}\right)\mathbf{\hat{b}}$.
We use this formal criterion to locate the reconnection regions in our simulation (red isocontour in Fig.~\ref{fig:4}{\bf d}).
The regions where the left-hand side of this expression substantially deviates from zero surround the Z-pinches, their locations correlate well with the regions of high electron dissipation $D_e$.

As an additional indicator of magnetic reconnection sites we use the deviation of the electron speed from the drift speed
$\mathbf{\upsilon}_{e,\perp} - {\left[\mathbf{E} \times \mathbf{B} \right]}/{B^2}$.
Where this difference is large, the electrons are decoupled, and the topology of magnetic field is likely to change, shown by green isocontour in Fig.~\ref{fig:4}{\bf d}.
More electrons are decoupled where the plasma is hotter (compare with Fig.~\ref{fig:4}{\bf b}).
Magnetic reconnection in pinches generates secondary oscillations with wavelength $\sim d_i$, noticeable between the adjacent pinches.
None of the considered reconnection indicators has substantial value in the vicinity of X-points, all non-ideal effects are associated with pinches.

\section{Summary}
We have performed a 3D simulation of relaxation of a non-equilibrium collisionless plasma.
The initial magnetic field configuration is such that non-spiral null-points are present in the simulation domain together with series of O-type null-points forming null-lines.
Due to initial pressure imbalance, the simulation starts with rapid relaxation, during which strong currents establish in the null-lines, creating Z-pinches with radii of the ion skin depth.
Almost half of the magnetic energy is released in this phase, also accelerating particles to suprathermal speeds.
When the pressure disbalance is compensated, stationary volumetric reconnection is driven primarily by the kinking of the Z-pinches that forms mixed-polarity fields inside and around the current channels.

The magnetic energy spectrum obtains its characteristic shape with two breaks (at large scales, and at electron scales) during a few ion gyration times.
The slope of the spectrum between ion and electron scales is very steep, with exponent $\gamma\approx-5$, suggesting the primarily one-dimensional dissipation processes.
Which conforms with the reconnection driven by pinches. 
In the phase of the steady reconnection, the dissipation rate of magnetic energy is $1.5$\% per ion gyro period. 
Most of the dissipated magnetic energy goes to heat the plasma, and does not substantially change the high-energy tails in particle distributions.

If observations suggest the large amount of spiral null-points in certain space plasmas, e.g., in the magnetosheath or in the solar wind, where the currents are also measured,
then the structures similar to pinches may form in these regions.
Magnetic reconnection driven by instabilities developed in these structures may be the dominating mechanism of magnetic energy dissipation.

Authors are thankful to Yu. Khotyaintsev and A. Vaivads for useful discussions.
This research has received funding from the European Commission's FP7 Program with the grant agreement SWIFF (project 2633430, swiff.eu). 
The simulations were conducted on the computational resources provided by the PRACE Tier-0 project 2011050747 (Curie supercomputer).

\bibliographystyle{jpp}
\bibliography{3dnull}

\end{document}